# Synergizing Self-Regulation and Artificial-Intelligence Literacy Towards Future Human-AI Integrative Learning

**Long (Jim) Zhang (u3008540@connect.hku.hk)**
Faculty of Education, The University of Hong Kong
Pokfulam, Hong Kong SAR

**Shijun (Cindy) Chen (shijunchen@connect.hku.hk)**
Faculty of Education, The University of Hong Kong
Pokfulam, Hong Kong SAR

## Abstract

Self-regulated learning (SRL) and Artificial-Intelligence (AI) literacy are becoming key competencies for successful human-AI interactive learning, vital to future education. However, despite their importance, students face imbalanced and underdeveloped SRL and AI literacy capabilities, inhibiting effective using AI for learning. This study analyzed data from 1,704 Chinese undergraduates using clustering methods to uncover four learner groups reflecting developing process—Potential, Development, Master, and AI-Inclined—characterized by varying SRL and AI literacy differentiation. Results highlight obvious disparities in SRL and AI literacy synchronization, with the Master Group achieving balanced development and critical AI-using for SRL, while AI-Inclined Group demonstrate over-reliance on AI and poor SRL application. The Potential Group showed a close mutual promotion trend between SRL and AI literacy, while the Development Group showed a discrete correlation. Resources and instructional guidance support emerged as key factors affecting these differentiations. To translate students to master SRL-AI literacy level and progress within it, the study proposes differentiated support strategies and suggestions. Synergizing SRL and AI literacy growth is the core of development, ensuring equitable and advanced human-centered interactive learning models for future human-AI integrating.

## Main

Generative Artificial Intelligence (GenAI), is reshaping human learning, as discovered by Yan et al. (2024). It brings personalized learning process by powerful adaptive support (e.g., Borah et al., 2024). With the deepening of AI integration into human learning, human-AI interactive learning has the potential to turn to human-AI integrated symbiotic (integrated) learning. For example, AI contributes valuable knowledge and capabilities to the learning process, and students, in turn, can adjust and optimize AI functionalities to better meet project requirements (human-AI symbiotic learning, Zheng et al., 2024). Although human-AI interactive learning is promising, human may not in the right path of learning. For example, humans are given the illusion that their knowledge is accumulating with the abundance of accessible information, however, the unique human skills may not properly cultivated, such as self-regulated learning (Yan et al., 2024). This self-perception bias may be further amplified in the continuous interactive cycle of learning with AI (Glickman & Sharot, 2024), which ultimately causes learners to overlook self-improvement.

To maximize the effectiveness of AI usage in human competencies cultivation, AI literacy is proposed, accompanied with caution against autonomy loss from over-reliance (Darvishi et al., 2024; Yan et al., 2024). AIL, defined by Lintner (2024, p.1) as "the ability to understand, interact with, and critically evaluate AI systems and outputs," is vital in human-AI learning. For caution against autonomy loss, self-regulated learning (SRL) skills are beneficial for maintaining human-centered learning. Self-regulation, as conceptualized by Bandura (1986), helps mitigate biases by enabling learners to manage their emotions and behaviors, which is crucial in ensuring that learning remains aligned with human needs and values. SRL empowers learners to maintain control and adaptability, while also upholding ethical standards. This, in turn, prevents passivity and ensures that learners actively shape the interactive learning process. By doing so, SRL supports a human-centered decision-making (Markauskaite et al., 2022) approach that prioritizes the agency and ethical considerations of learners, even in the face of potential challenges from external influences.

Earlier literature presented with evidences that SRL and AIL interplay with each other during human-AI interactions. Firstly, self-regulation skills help equips learners to integrate AI (Lodge et al., 2023). SRL empowers students to effectively and responsibly use AI tools to learn. SRL's enhancing can help suit with rapid changes by AI integrating (re-skilling/up-skilling) and maintaining human-centered decision-making (Markauskaite et al., 2022). Therefore, the improvement of SRL can make students better integrate AI into learning, that is, improvement of AI literacy. Secondly, AI also benefit to SRL. For instance, Giannakos et al. (2024) asserted the significant potential of GenAI in promoting students' SRL. Similarly, studies conducted by Ng et al. (2024) and Kong & Yang (2024) successfully leveraged GenAI to foster students' development in SRL. Given these successes, AI literacy can enable students to better use AI to facilitate their SRL.

Given the importance and interplaying of SRL and AIL in human-AI interactive (or even integrative) learning, both should be synergize-developed. However, the current landscape is troubling. Most learners lack essential competencies, with some showing maladaptive patterns. Even top performers face challenges - a critical concern given AI's rapid integration into work, life, learning before adequate preparation.

In the survey in mainland China (n=1704), we identified four clusters. Potential Group (n=792), Development Group (n=512), Master Group (n=165), and AI Inclined Group (n=235), represented stages of students' SRL-AIL synergized development. The Potential Group has low SRL and AIL levels, the Master Group high levels, while the Development Group lies in between. The AI Inclined Group shows AIL levels similar to the Master Group but SRL levels aligned with the Development Group. Correlations between SRL and AIL differ: the Master Group shows critical human-AI learning, the Potential Group has a widespread positive correlation, while others reveal imbalanced and contradicted patterns. AI-Inclined Group should be noticed, which could suffer from over-reliance AI.

We found that resources (economic & technical) and instructional support were the factors in the differentiation. Resource support has the significant impact on the Potential Group, while the Development Group and the AI-inclined Group should be provided synergizing SRL-AIL training (rather single training). However, the Master Group is not the final of development, and they still need more in-depth guidance to continue to progress. Thus, educators should not only just support and guidance for students to develop to the master level, but also keep theoretical and technical exploration for the master level, towards the advanced human-AI interactive or even the integrated learning.

## Results

### Clustering Results and Naming

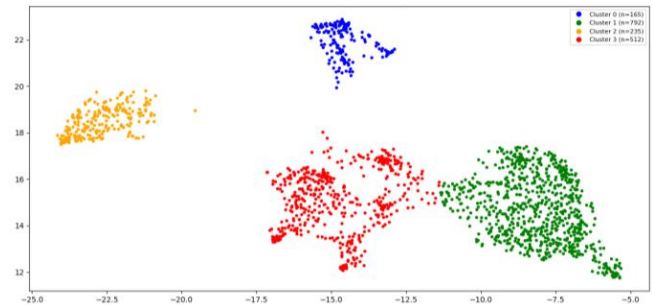

Figure 1. UMAP+K-Means clustering SRL and AIL

Uniform Manifold Approximation and Projection (UMAP) is a popular dimensionality reduction technique that preserves both local and global data structure characteristics. K-Means is a commonly used clustering algorithm that minimizes intra-cluster variance by assigning data points to k clusters. By integrating UMAP and K-Means, we achieved optimal clustering results, identifying four clusters as shown in the figure. Optimized UMAP parameters are {n_components=2, n_neighbors=15, min_dist=0.2}, and K-Means parameters are {n_clusters=4}. Ultimately, the silhouette score of the clustering is 0.668, indicating that data points are tightly grouped within their clusters and clearly separated from others. According to Kaufman and Rousseeuw (1990), a silhouette score of 0.668, approaching 0.7, falls within the transition range from reasonable to strong clustering structure.

Cluster 0 shows high SRL and AIL, ideal for educational interventions, and leads in AI-Evaluation due to superior SRL. Cluster 1 has low SRL and AIL, needing targeted support. Cluster 2 and Cluster 3 both show moderate SRL but differ in AIL: Cluster 2 excels in AIL compared to Cluster 3. Cluster 2 outpaces Cluster 3 in AIL and slightly in monitoring. Cluster 3 surpasses Cluster 1 in SRL but barely in AIL. Cluster 0 (Master Group, MG) (n=165) excels in SRL and AI. Cluster 1 (Potential Group, PG) (n=792) ranks lowest. Cluster 2 (AI-Inclined Group, AG) (n=235) matches SRL with Cluster 1 but rivals Cluster 0 in AI. Cluster 3 (Development Group, DG) (n=512) aligns with Cluster 2 in SRL but falls behind in AIL.

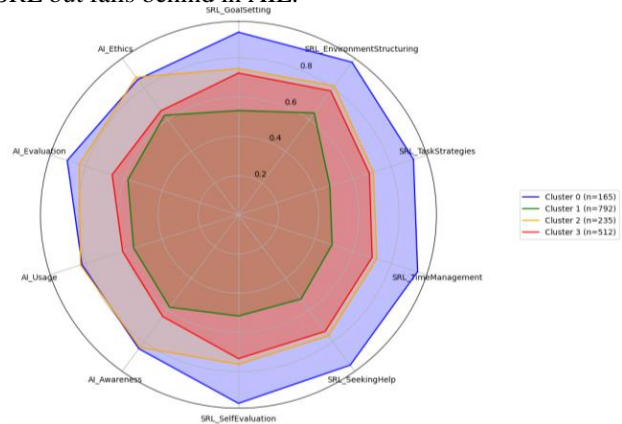

Figure 2. Radar Chart for clusters' SRL&AIL sub-dimensions

### Correlation Analysis between SRL and AIL

Used Spearman's rank correlation analysis, the relationships between SRL skills and AIL dimensions were examined (*stands for p value <.05; ** stands for <.01). MG with high SRL levels exhibits strong planning and management skills, effectively utilizing AI tools to achieve human-AI synergy in learning. MG shown well understanding of AI tools and SRL abilities, representing critical human-AI learning. They can use AI critically in self-regulated human-AI interactive learning processes. As positively correlates between SRL and AI evaluation (r=0.192*), while AIL correlates with self-evaluation (r=0.159*). They consider the correlation between self-evaluation and evaluation of AI output. More detailed evidence as AI evaluation correlates positively with SRL's goal setting (r=0.217**), self-evaluation(r=0.319**), and help-seeking (r=0.235*), while self-evaluation links to AI usage (r=0.172*). These indicate that students will use AI outputs critically when performing these SRL sub-dimensional behaviors, and will also conduct self-evaluation when using AI, during the learning process. However, MG still requires further guidance and intervention of SRL-AIL synergizing training. They consider AI ethics, but it limits their time management / task strategies. As the significant negative correlation between AI ethical and task strategy (r = -0.236**) and time management (r = -0.215**).

PG shown widespread positive correlations. They are in a phase of natural development. Reinforcing self-evaluation and environmental structuring significantly enhances AIL (overall SRL-AI correlation: r=0.188**). For more detailed evidence, improvements in SRL (overall) exert a significant positive impact on the sub-dimensions of AIL: AI awareness

(r=0.135**), AI usage (r=0.151**), AI evaluation (r=0.141**), and ethics (r=0.192**). While the AIL (overall) also benefit whit SRL's sub-dimensions: goal setting (r=0.084*), environmental structuring (r=0.251**), help-seeking (r=0.203**), and self-evaluation (r=0.117**). Notably, task strategies and time management showed barely non-significant correlation with overall AI literacy or most of its sub-dimensions. Results of inexistent significant negative correlation suggest that PG development is natural and not strong needs of SRL-AIL synergizing training intervention.

DG has demonstrated an immature critical use of AI's ability to self-regulated human-AI interactive learning. AI evaluations show consistently positive correlations with three subprocesses of SRL: environmental structure (r=0.106*), help-seeking (r=0.150**), and self-evaluation (r=0.126**). However, students' self-evaluation abilities lack significant correlations with other AI literacy dimensions (i.e. awareness, use, ethics). In contrast, help-seeking has a significant positive correlation with overall AI literacy (r=0.089*). This indicates that, although progressing towards the MG, these students have not yet developed the capacity of using AI in a critical and evaluative approach. They need SRL-AIL synergizing training toward MG level. Below significant negative correlations also highlight the needs. AI awareness shows significant negative correlations with overall SRL (r=-0.103*), goal setting (r=-0.088*), and environmental structure (r=-0.051*); task strategy capacity shows significant negative correlations with overall AI literacy (r=-0.116**), AI evaluation (r=-0.116**), and AI ethics (r=-0.101**); and time management shows a significant negative correlation with AI ethics literacy (r=-0.170**).

AG completely deviates from the path of development phase towards MG. They lack critically AI using in learning but over-reliance. As AI awareness significantly positively correlates with environment structure (r=0.207**) and help-seeking (r=0.210**). While AG's AI evaluation is not significantly related to environment structuring, help-seeking, and self-evaluation subprocesses in SRL; self-evaluation also not significantly related all of AIL and its sub-dimensions. Moreover, task strategies show significant negative correlations with AIL overall (r=-0.169**), AI usage (r=-0.148**), AI evaluation (r=-0.140*), and AI ethics (r=-0.282**). While AI usage shows significant negative correlations with students' goal setting (r=-0.141*), task strategies, and time management (r=-0.131*), further emphasizing the threats of AI over-reliance. Above of this over-reliance, given AG's general negative and non-significant correlation of SRL-AIL, we strongly suggest synergizing training. In particular, the combination of AI evaluation and self-evaluation is used to the self-regulated human-AI interactive learning.

## Analysis of External Factors

### Gaps in External Resource Support
We examined external support using GDP (gross domestic product), disposable income, and tech coverage (phones, computers). External factors significantly impacted the potential group more than others. The Mann-Whitney U test showed a significant difference between the PG and others (all p < 0.01), while no significant differences were found among the other three groups.

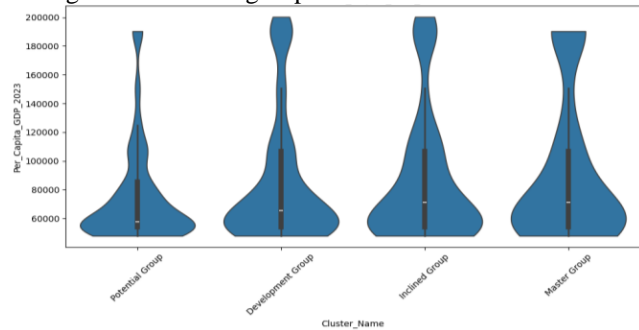

Figure 3. GDP per capita differs between the groups

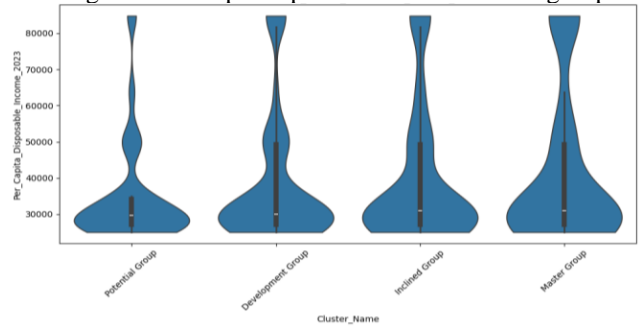

Figure 4. Disposable income per capita differs between the groups

Further comparison using box plots revealed the noticeable disparity occurring in the central portion (Fig.3 & Fig.4), which underscores the importance of economic resource support for student development. Regarding the equipment indicators, it was observed that, compared to other groups, the PG exhibited highest mobile phone coverage but the lowest computer coverage (Fig.5 & Fig.6). This highlights the significance of the type of equipment support for student development with synergizing training. Additionally, this disparity reveals underlying educational inequities (Cf. Yan et al., 2024). To address these issues, it is necessary to allocate more economic and equipment for PG.

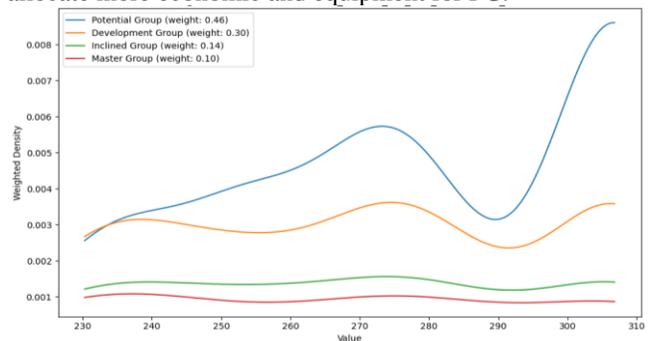

Figure 5. Mobile phones penetration between groups

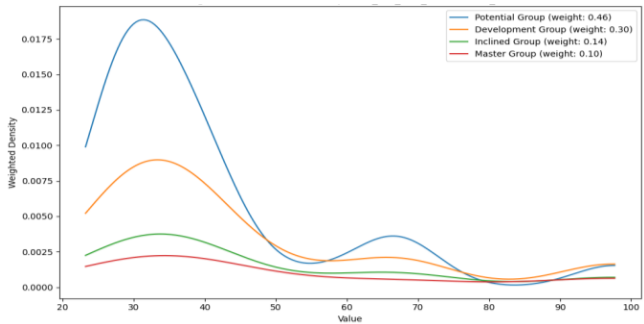

Figure 6. Computers penetration between groups

**Gaps in Instructional Guidance Support**

In the analysis of instructional guidance, we observed a pronounced evolutionary pattern in the training models for personal capability development. Statistical results (all p <0.01 by Chi-Square) further confirmed the core statement for synergizing SRL-AIL training as instructional support rather than developing only one of them. Synergizing training support proportions are increased from PG to MG (Synergizing, PG，19.07%; DG, 31.45%; AG, 38.30%; MG, 47.88%). While only SRL training proportions are decreased and only AIL training proportions are similar (SRL only, PG，48.36%; DG，48.24%; AG，45.11%; MG, 34.55%; AIL only, PG, 4.17%; DG，3.32%; AG，3.40%; MG，4.85%).

These findings highlighted again that the significant personal capability growth does not rely on single-mode training but requires synergistic SRL and AL training to exert promising learning effects. This cross-domain and integrated training approach provides a critical pathway for transforming individuals from potentials to experts. Meanwhile it can avoid the over-reliance on AI during learning process and presents a systematic evolutionary mechanism for personal growth.

## Discussion

**Mechanisms of Student Group Formation**

In the Figure 7, we integrated the elements of AI into a theorized SRL model developed by Zimmerman (1986) to further clarify our results and future research and practice implications. It demonstrates how SRL and AIL interplay with each other in human-AI integrations as the aforementioned results show.

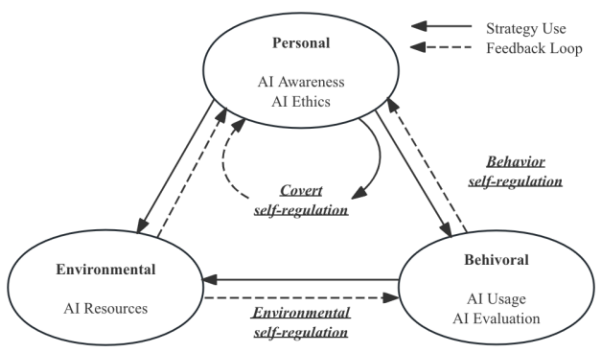

Figure 7. A Social Cognitive View of SRL-AIL adapted from Zimmerman (1986)

In this process (Fig. 7), AI awareness and ethics are personal factors, while AI use and evaluation are behavioral factors, which can collectively influence students' self regulation strategy usage. For example, AI evaluation influenced their choice of different AI resources (e.g., Zheng et al., 2024), which diff of AI evaluation literacy reflected environmental self-regulation (choice of different AI resources). As counter, the enhancement of SRL has great potential to the AIL improvement in a feedback loop. For example, students could optimize their AI prompts and develop academic achievements through multiple human-AI interactions (e.g., Wan & Chen, 2024), which is taken as one type of behavioral self-regulation in our research. From the above, we can conclude that SRL and AIL are reciprocal in interplaying with each other.

At the environmental dimension of using AI in SRL, AI resources (e.g. external opportunities to access technology and educational support) as significant factors to cluster different learning groups and personal learning inclination. Over time, these varying inclinations broaden gaps among groups. Such as lack of sufficient synergized instructional support, somebody deviate from the development path, resulting in over-reliance on AI. To guide learners toward and keep progressing in master level, we propose differentiated support strategies in the following section.

**Differentiated Support and Guidance Strategies**

For PG, Their SRL-AIL is at a rudimentary level and grows naturally without the need for strict intervention. They are higher encouraged and provided more opportunities to access human-AI interactive learning. It is essential to focus on economic and equipment support. Notably, computers rather than mobile phones are better equipment support to synergized develop SRL-AIL. Using computers for synergized SRL-AIL development, it is easier for students and AI to collaborate (such as writing task while AI collaboration), thus better implementing human-AI interactive learning (e.g., Zheng et al., 2024). By contrast, with a mobile phone, students are talking directly to the AI (such as asking question), rather than working collaboratively on the task. Furthermore, mobile phones are predominantly seen as entertainment devices that are more prone to interruptions, making them a less effective learning environment structure (e.g. Deng, 2020).

For DG, we hereby advocate to provide personalized synergizing training intervention so as to nurture both self-evaluation capabilities and AI literacy. With self-evaluation, students should learn to evaluate AI-provided content critically, integrate it into their learning and use it ethically. Teachers can scaffold in pedagogical practices on how to combine self-evaluation and AI evaluation.

For AG, interventions targeting SRL-AIL synergizing development are needed. First, teachers should guide students to set holistic academic goals that extend beyond grades, emphasizing the development of professional competencies such as creativity, problem-solving abilities, and self-regulation (Cf. Yan et al., 2024). Secondly, Teachers need to help students develop their awareness and ethic of properly using AI and the risks of AI over-reliance. (Cf., Ng et al., 2021). Finally, students are suggested to take self-evaluation as a starting point in their self-regulation if they aim to transit to the master level. Teachers can conduct it in simulated tasks (e.g., Zheng et al., 2024), so that students understand that self-regulating learning in human-AI interaction can enable them to achieve greater output (grades) and potential benefits (creative thinking, self-regulating ability, etc.). For example, AI usage reflection workshop can be conducted for students to reflect how do they use AI and evaluate if they and their peers have used it properly. Such reflections offer great opportunities to raise students AI awareness and ethical norms.

We need to explore future technical and theoretical developments to further guide the MG group development. When students reach advanced Master levels SRL-AIL synergizing development, they showed the characteristics of critical human-AI interactive self-regulated learning. AI ethical literacy reduce and limit the development of choosing learning strategies and time management. These limitations appears when students enter the developmental process (including DG & AG). One possible explanation is that when students are engaged in self-regulated learning, they may be using AI tools to help them plan and manage their time, monitor progress, or adjust learning strategies. However, as their AI ethical literacy improves, they may become more concerned about the ethical issues of these tools, such as data privacy, algorithmic bias, and so on. As a result, they choose fewer task strategies and spend more time thinking about them. Therefore, educators should explore further technical and theoretical exploration to these limitations: improving the learning process while maintaining the ethic of "AI for social good" (e.g., Novitsky, 2024; Elantheraiyan et al., 2024).

**Implications**

In the evolving landscape of SRL-AIL, concerns surrounding AI often inadvertently constrain the development of learning behaviors and strategies. This limitation is particularly pronounced among individuals with more advanced SRL-AIL competencies. To address this challenge, we propose an integrative human–AI learning process model. The model's principal contribution lies in its paradigm shift: rather than treating humans and AI as separate entities, it conceptualizes them as a symbiotic whole engaged in a unified learning process. This reconceptualization holds promise for surmounting existing constraints in SRL-AIL and fostering further development.

Regarding the progression from human-computer interaction to symbiotic integration, existing studies provide guidance. In terms of how the progression works, Sowa et al. (2021) propose a four-phase progression from interaction to deep fusion, culminating in a human-computer hybrid state. Zhang et al. (2022) refine their process from trust-building to human-computer symbiosis, emphasizing collaborative advantage realization to maintain strengths of both humans and AI. This evolution is dynamic and ongoing, requiring constant adaptation and alignment by both humans and AI. To facilitate better human-computer collaboration, Almeida and Senapati (2024) suggested the standards, include model interpretability, fostering trust, understanding decision processes, and ensuring alignment with human values, with the core principle ensuring human primacy in key decision-making. Integrating these processes and standards into learning promotes the development of human-computer collaborative learning systems (Cf. Yan et al., 2024). Based on above, we explored a learning process framework during Human-AI integrated (Fig. 8) from a constructivist perspective, viewing humans and AI as a unified entity interacting with the external world.

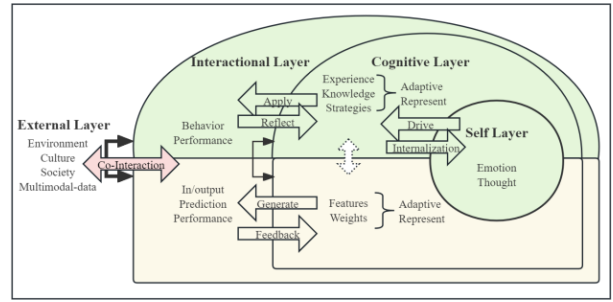

Figure 8. The Human- AI Integrative Learning Process

In this framework, humans and AI are regarded as a joint learning entity (referred as "Cyberlearners") that collectively engage with external elements such as learning environments, sociocultural contexts, and multimodal data. Through these co-interactions, both human and AI components learn in tandem. For instance, during specific tasks (e.g., Report Design , Zheng et al., 2024), AI, serving as a source of knowledge affordances, augments the learner's capacities, while learners iteratively refine AI through continual use and feedback, thereby enhancing task alignment (Zheng et al., 2024). In this co-interaction with External Layer process, the human learning components are depicted in green, while the AI computational learning elements are in yellow. At the Interactional Layer, humans engage with the external layer through behaviors and performances, whereas AI interacts via inputs/outputs, predictions, and other functionalities. Through this interaction, the Cognitive Layer of the Human-AI expands, encompassing human experiences, knowledge, metacognition, and AI features, weights, etc. Importantly, the human-centric approach necessitates sustaining human emotions, thoughts, and considerations of AI on a Self Layer.

Implementing this human–AI integrative learning approach can transcend the constraints of SRL-AIL by reconfiguring AI from an external tool to a symbiotic partner within the learning process. It's different from that learners must actively invoke AI, which remains an external entity. Considering AI as the symbiosis of learners further mitigates concerns about privacy and bias, allowing learners to remain focused on meaningful engagement. Meanwhile, differently with evaluating AI's ethical implications, the Cyberlearners model aligns AI's ethical frameworks with those of the learner, obviating the need for separate assessments. From above, this integrative framework effectively addresses privacy, bias, and ethical dilemmas at their source.

We encourage future empirical studies to validate and expand this model, for example, through the tech-development of symbiotic, self-regulation guidance for Cyberlearners, and updated assessment standard for human-AI integrated learning. We believe that through continuous technical and theoretical exploration, human-AI integrative learning will further expand the boundaries of human capabilities and keep progressing synergized development of SRL-AIL. As Jarrahi (2018) observes, AI should bolster human capacities and catalyze collaborative progress between humans and machines. The role of AI in the future of learning should not be underestimated.

# Method

## Instrumentation

With the growing popularity of online learning, Barnard et al. (2009) developed the OSLQ (Online Self-Regulated Learning Questionnaire), more suited to online environments. Given that AI literacy is an advanced form of digital literacy (Yang, 2022), this study employs the OSLQ as the SRL measurement tool. AIL measurement follows the recommendations of Lintner (2024), utilizing the AILS (AI Literacy Scale) (Wang et al., 2022), which has proven to be an excellent indicator for assessing AIL in the general population.

## Participants and Data Collection

This study adopts an online survey method, recognized for its efficiency in cross-sectional studies in China (Liu et al., 2023). To ensure the representativeness of the data, the survey covered 34 universities distributed across the eastern (6 provinces, 13 universities), central (5 provinces, 9 universities), and western regions (6 provinces, 12 universities).

In late 2024, the AILS and OSLQ questionnaires were distributed electronically. Participation was voluntary, with no incentives or academic penalties. Quality control measures excluded extremely fast responses (Liu et al., 2023), patterned replies (Hair et al., 2009), and outliers beyond ±3 standard deviations (Field, 2013). The final dataset included 1,704 valid responses from 448 males and 1,256 females across 34 universities. Respondents were mainly 18–24-year-old undergraduates (95.8% and 92.9%, respectively), spanning disciplines such as education, science, arts, and economics and management. The methodology aligns with Liu et al. (2023), confirming the efficacy of online surveys in reaching diverse student groups.

The questionnaires assessed individual SRL, AIL, and environmental factors based on social cognitive learning theory, focusing on instructional and environmental support. We asked students whether they had received SRL or AIL training. The teaching support was measured through their self-reported outcomes. Indicators for environmental support included provincial 2023 per capita GDP (Wang, 2023), disposable income (Zhao et al., 2023), and 2022 mobile-phone/computer (Mendoza, 2014) ownership per 100 households, reflecting the role of economic and equipment levels in education.

## Data Analysis

The data analysis process included clustering, naming, internal analysis, and external factor exploration. Clustering aimed to identify groups with similar behaviors and responses. Naming analyzed and characterized these groups. Internal analysis examined differences and correlations between students' SRL and AIL. External factor analysis explored instructional support and environmental indicators, explaining group formation and proposing personalized support.

After standardizing the questionnaire indices, we categorized learners using clustering algorithms. The questionnaire data included 24 SRL items and 12 AI items, resulting in 36 dimensions. Guided by Allaoui et al. (2020), to improve clustering computational efficiency, UMAP was used for dimensionality reduction. Upon comparison, the K-means algorithm was adopted, and parameters were optimized using the Optuna tool (e.g., Hadianti & Kodri, 2023). The silhouette coefficient confirmed the best performance of K-means, identifying four clusters.

We named and analyzed the four clusters: based on the mean standardized scores of SRL and AI, and conducted in-depth comparisons across six SRL subdimensions and four AI subdimensions. The four clusters were: high performers, AI-oriented individuals, SRL-oriented individuals, and low performers. As non-normal distribution, Spearman's rank correlation analysis explored the relationship between SRL and AI (Sedgwick, 2014).

Further analysis revealed significant differences in external factors across clusters. Geographic and economic equipment indicator data were analyzed using weighted Gaussian kernel density estimation to address sample size variance. Differences in economic and equipment levels were analyzed using the Mann-Whitney U test, identifying specific environmental characteristics of each cluster, due to non-normal distributed (MacFarland et al., 2016). Descriptive statistics used for checking differences in teaching guidance support with independence check by Chi-square.